\documentclass{PoS}

\title{Microscopic studies on nuclear spin-isospin properties---a personal perspective on covariant density functional theory}

\ShortTitle{Microscopic studies on nuclear spin-isospin properties in CDFT}

\author{\speaker{Haozhao Liang}\thanks{This series of works was partially supported by the RIKEN iTHES and iTHEMS projects, the Grant-in-Aid for JSPS Fellows under Grant No. 24-02201, and the National Natural Science Foundation
of China under Grant No. 11105006.}\\
        RIKEN Nishina Center, Wako 351-0198, Japan\\
        Graduate School of Science, the University of Tokyo, Tokyo 113-0033, Japan\\
        E-mail: \email{haozhao.liang@riken.jp}}

\abstract{Spin and isospin are essential degrees of freedom in nuclear systems, and the relevant studies on their properties play important roles not only in nuclear physics but also in nuclear astrophysics, particle physics, and so on.
In this presentation for the IUPAP Young Scientist Prize 2016, I would like to introduce the microscopic studies on nuclear spin-isospin properties in the framework of covariant density functional theory (DFT), by taking a few works that I have been joining in as examples.
It is seen that the covariant scheme plays an important role in describing the spin properties in a consistent way, such as the spin-orbit splitting, the pseudospin symmetry, etc.
Meanwhile, the Fock terms play important roles in describing the isospin properties fully self-consistently, such as the Gamow-Teller and spin-dipole resonances, the isospin-symmetry-breaking corrections to nuclear superallowed $\beta$ transitions, etc.
To connect the covariant DFT to more fundamental theories, an \textit{ab initio} relativistic Brueckner-Hartree-Fock theory for finite nuclei is introduced, and a personal perspective for the coming decade is also illustrated.
}

\FullConference{The 26th International Nuclear Physics Conference\\
		11-16 September, 2016\\
		Adelaide, Australia}

\begin{document}

\section{Introduction}

Atomic nucleus is quite a unique quantum-mechanical many-body system, in which three fundamental interactions out of four---the strong, weak, and electromagnetic interactions---interplay each other in a large range of time and energy scales with the co-existence of single-particle and collective characteristics.
It holds profound properties as a quantum system, a many-body system, a finite system, as well as an open system.
One of the best examples to identify these features is the halo structure in neutron drip-line nuclei.
For more than a half century, we learnt in textbooks that the size of a nucleus would be essentially proportional to its mass number because of the saturation property of nuclear force.
However, the nature surprised us in 1985.
It was found that the size of $^{11}$Li is almost the same as that of $^{208}$Pb due to its exotic neutron-halo structure \cite{Tanihata1985}.
From then on, our knowledge in nuclear physics is rapidly expanding together with the constructions and upgrades of the radioactive-ion-beam facilities all around the world.

To study such a complicated system, of most importance is to catch the essential degrees of freedom, for example, the spin and isospin degrees of freedom.
On one hand, the spin degree of freedom is one of the cornerstones in nuclear physics.
Only after the strong spin-orbit interaction was taken into account, the traditional magic numbers in stable nuclei were understood.
Meanwhile, by examining the single-particle spectra, Hecht and Adler \cite{Hecht1969} and Arima, Harvey, and Shimizu \cite{Arima1969} found the near degeneracy between two single-particle states with quantum numbers $(n,\,l,\,j = l + 1/2)$ and $(n-1,\,l + 2,\,j = l + 3/2)$.
They introduced the concept of pseudospin symmetry to describe such an approximate degeneracy.
On the other hand, the isospin degree of freedom distinguishes protons and neutrons in nuclei.
The physics becomes much richer once there exist two different but similar kinds of fermions in one system.
Note that nuclei will never be self-bound if they are composed of only protons or neutrons.
It comes the concept of symmetry energy in nuclear equation of state, and this isospin property becomes one of the frontiers in nuclear physics and astrophysics, since it is crucial to understand why there are two-solar-mass neutron stars in our Universe.

Instead of investigating the nuclear spin and isospin properties separately, experimentally, one of the best probes to study these two essential degrees of freedom together is the so-called nuclear spin-isospin excitations \cite{Osterfeld1992}, such as the Gamow-Teller (GT) and spin-dipole (SD) resonances.
These excitations correspond to the transitions from an initial state of the nucleus $(N,\,Z)$ to the final states in its isobaric neighboring nuclei $(N-1,\,Z+1)$ and $(N+1,\,Z-1)$ in the isospin lowering and raising channels, respectively.
They can take place spontaneously in nature, like the well-known $\beta$ decays, or be induced by external fields in laboratory, like the charge-exchange reactions, e.g., $(p,\,n)$, $(n,\,p)$, $(^{3}{\rm He},\,t)$, which have been intensive studied, e.g., in RIKEN, RCNP, MSU, GSI, TRIUMF, CERN, etc.
That is because these excitations are important to understand the questions like ``\textit{What are the spin and isospin properties of nuclear force and nuclei?}'' ``\textit{Where and how does the rapid neutron-capture process (\textit{r}-process) happen?}'' ``\textit{Does Cabibbo-Kobayashi-Maskawa (CKM) matrix satisfy the unitary condition?}'' etc.
These are among the top questions in nuclear physics, nuclear astrophysics, and particle physics.
One of our main tasks is to understand these questions from the theoretical side.

\section{Covariant density functional theory}

For these studies, one of our favorite research tools is the nuclear density functional theory (DFT), in particular, its covariant (or the so-called relativistic) version \cite{Walecka1974}.
Its fundamental is the Kohn-Sham DFT and its scheme is the Yukawa meson-exchange picture.
The starting point of covariant DFT is an effective Lagrangian density with a typical form as
\begin{eqnarray}
    \mathcal{L}&=&\bar\psi\left[i\gamma^\mu\partial_\mu-M-g_\sigma \sigma-g_\omega \gamma^\mu \omega_\mu
        -g_\rho \gamma^\mu \vec\tau \cdot \vec \rho_\mu-\frac{f_\pi}{m_\pi}\gamma_5\gamma^\mu\partial_\mu\vec\pi\cdot\vec\tau
        -e\gamma^\mu\frac{1-\tau_3}{2}A_\mu\right]\psi\nonumber\\
    &&+\frac{1}{2}\partial^\mu\sigma\partial_\mu\sigma-\frac{1}{2}m^2_\sigma\sigma^2
        -\frac{1}{4}\Omega^{\mu\nu}\Omega_{\mu\nu}+\frac{1}{2}m^2_\omega\omega^\mu\omega_\mu
        -\frac{1}{4}\vec{R}^{\mu\nu}\cdot\vec{R}_{\mu\nu}+\frac{1}{2}m^2_\rho\vec\rho^\mu\cdot\vec\rho_\mu\nonumber\\
    &&+\frac{1}{2}\partial_\mu\vec\pi\cdot\partial^\mu\vec\pi-\frac{1}{2}m^2_\pi\vec\pi\cdot\vec\pi
        -\frac{1}{4}F^{\mu\nu}F_{\mu\nu}\,,
\end{eqnarray}
in which nucleons are described as Dirac spinors that interact each other via the exchanges of the $\sigma$, $\omega$, $\rho$, and $\pi$-mesons and photons.
Some details can be found in recent Reviews \cite{Vretenar2005, Meng2006, Paar2007, Niksic2011, Liang2015, Meng2015} and the references therein.

One of the key reasons for choosing DFT is because it is applicable to almost the whole nuclear chart, and not only for the ground states but also for the excited states.
In particular, for choosing the covariant version, a dream here is starting with an effective Lagrangian we are eventually able to connect the nuclear DFT to more fundamental theories, such as QCD at low energy.
Even before that, from the practical point of view, by using the Dirac equation as the nucleons' equation of motion instead of the Schr\"odinger one, the spin degree of freedom and a large part of the three-body effect can be taken into account in a consistent way.
Moreover, the Lorentz covariant symmetry holds a unified description of the time-even and time-odd components of the energy density functionals.

Therefore, a great effort has been devoted to the nuclear covariant DFT since the Walecka model proposed in the 1970s \cite{Walecka1974}.
At the beginning of the year 2016, a kind of summary for the cutting-edges in this field was published in the latest volume of \textit{International Review of Nuclear Physics}---\textit{Relativistic Density Functional for Nuclear Structure} \cite{Meng2016}.
Some of the highlights have been covered by Professor S.G. Zhou in his plenary talk in this conference.
Here I show a few works that I have been joining in as examples.

An example is for the pseudospin symmetry.
It is found that although it is deeply hidden in the original Hamiltonian, the origin of pseudospin symmetry can be traced in its supersymmetric partner Hamiltonian.
For the first time, the pseudospin-orbit splitting can be understood in an explicit and quantitative way  \cite{Liang2013, Shen2013, Liang2015}.

Another example is for the single-particle resonances.
In order to probe the single-particle resonances in the relativistic scheme, a method for solving the Dirac equation in the complex momentum space was developed recently.
It has been found that this method is not only very effective for the narrow resonances, but also can be reliably applied to the broad resonances \cite{Li2016, Fang2016}.
This is potentially crucial for understanding some exotic properties of halo nuclei.

One more example is for the nuclear rotation.
By developing the two-dimensional tilted-axis cranking model with the covariant DFT, the shears mechanism in the nuclear magnetic rotation and the two-shears-like mechanism in the anti-magnetic rotation can be described and understood in a self-consistent and microscopic way \cite{Zhao2011PLB, Zhao2011PRL, Zhao2012}.

In short, it is seen that the nuclear covariant DFT achieves a great success in describing and understanding various kinds of nuclear ground-state and excited-state properties.
Nevertheless, for simplicity, in most of versions of covariant DFT, only the local Hartree terms are kept, and the involved non-local Fock terms are neglected.
As will be shown below, these Hartree-only approaches show their limitations in some important features, like the properties in the spin-isospin channel and the tensor effects of nuclear interaction \cite{Liang2012}.
Back to my Ph.D. study, one of our key tasks was to develop a covariant DFT with both Hartree and Fock terms.

\section{Nuclear spin-isospin excitations}

The most typical nuclear spin-isospin excitation is the so-called Gamow-Teller resonances.
Their typical response functions are composed of a low-energy peak (around the $\beta$-decay window in unstable nuclei) and a high-energy broad peak as a giant resonance.
The absolute positions of these peaks tell us the isospin properties of the target nucleus, and the relative distance between these two peaks tells us its spin properties.

\begin{figure}\centering
    \includegraphics[width=.458\textwidth]{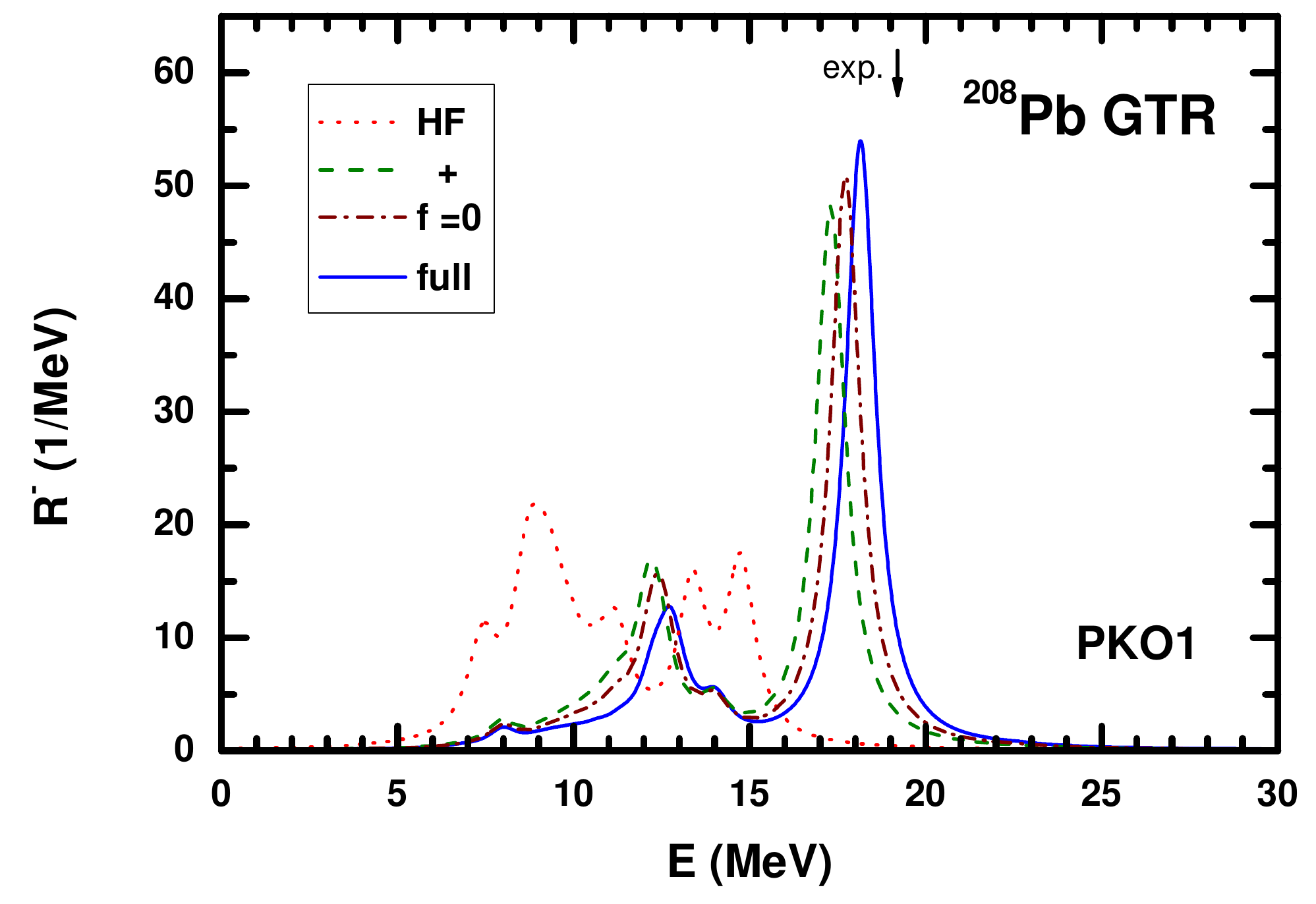}
    \hspace{1em}
    \includegraphics[width=.45\textwidth]{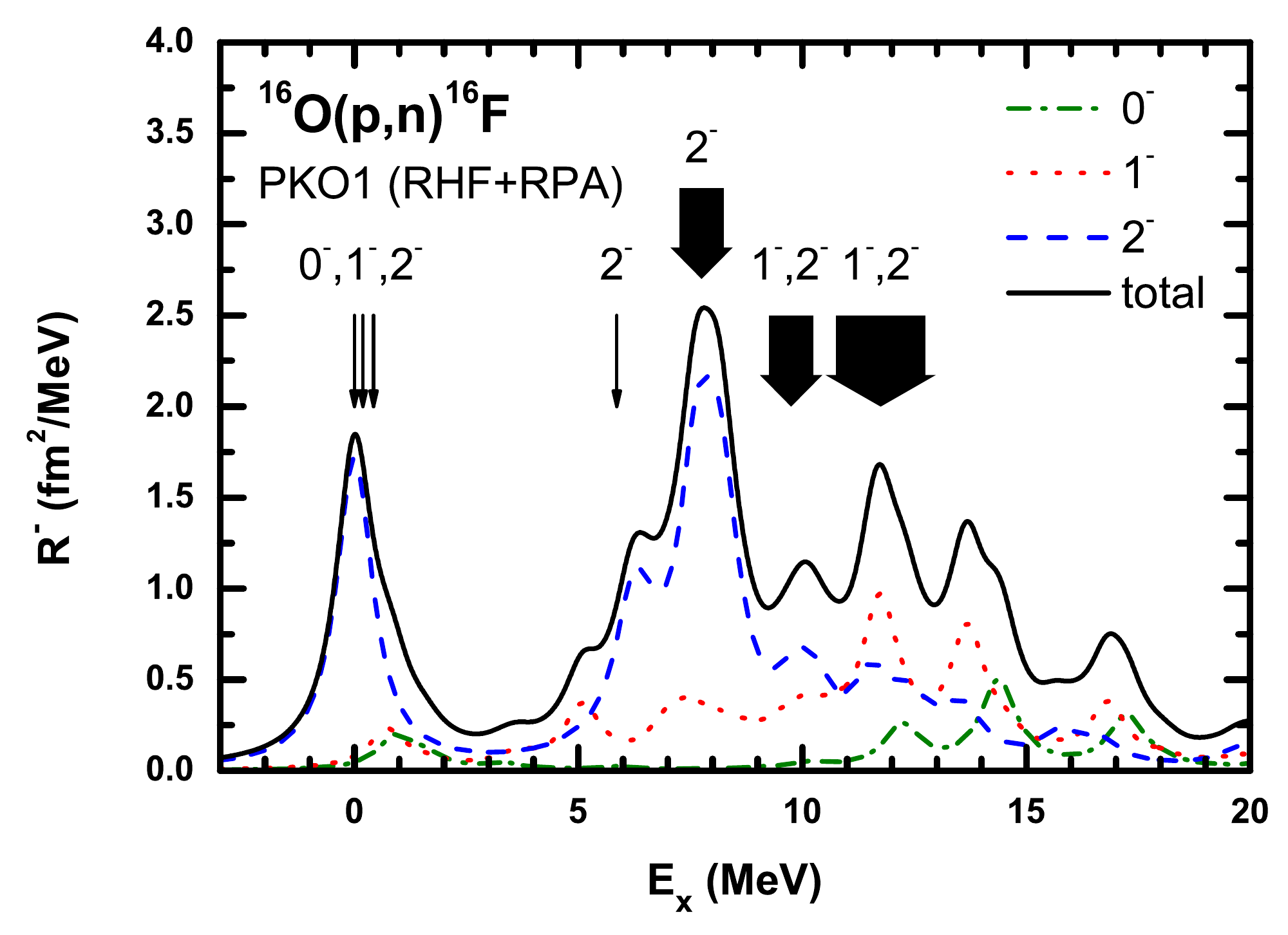}
    \caption{Strength distributions of the Gamow-Teller resonances in $^{208}$Pb (left) and the spin-dipole excitations in $^{16}$O (right) by the self-consistent RHF+RPA approach. Taken from Refs.~\cite{Liang2008, Liang2012O}.}
    \label{Fig1}
\end{figure}

As shown in the left panel of Fig.~\ref{Fig1}, the GT resonances in $^{208}$Pb can be reproduced by the random-phase-approximation calculations based on the relativistic Hartree-Fock theory (denoted as RHF+RPA).
With both Hartree and Fock terms, it was achieved for the first time in the relativistic scheme in a fully self-consistent way \cite{Liang2008}.
To discover the physics behind, we switched on and off individual residual interactions provided by each meson.
It has been found that in the present self-consistent scheme the isoscalar mesons play the essential roles via the Fock terms.
By using the Fierz transformation, one can also understand the reason why the previous calculations with only Hartree terms work with a free parameter $g'$ being around $0.6$ \cite{Liang2012}.

The predictive power of this RHF+RPA approach can be further examined by more delicate spin-isospin excitations, e.g., the spin-dipole excitations, which have three different components instead of one.
In the right panel of Fig.~\ref{Fig1}, the SD excitations in $^{16}$O are shown as an example.
Note that the three components could not be distinguished experimentally until an experiment with the high-quality
polarized proton beam performed in RCNP in 2011 \cite{Wakasa2011}.
It is seen that this delicate excitation structure can be described by the fully self-consistent RHF+RPA approach in details \cite{Liang2012O}.
This is not only important to understand the nuclear structure, but also important for some interdisciplinary studies, such as the $\beta$-decay have-lives of neutron-rich nuclei for the \textit{r}-process \cite{Niu2013} and the $\beta^+$ decays and electron captures of proton-rich nuclei \cite{Niu2013+}.
For example, we pointed out a remarkable speeding up of \textit{r}-matter flow, which leads to the enhanced \textit{r}-process abundances of the elements with $A \geq 140$ \cite{Niu2013}.

\section{Unitarity test of CKM matrix}

Another interesting application of RHF+RPA approach is for evaluating the isospin-symmetry-breaking corrections $\delta_c$ to nuclear superallowed $\beta$ transitions \cite{Liang2009}, which is shown to be crucial for the unitarity test of the CKM matrix.

The CKM matrix links between the quark eigenstates of week interaction and its eigenstates of mass.
In the latest versions of \textit{Review of Particle Physics} by Particle Data Group (PDG) \cite{PDG}, it is shown that nowadays the most precise unitarity test for the CKM matrix comes from the square sum of its first-row elements $V^2_{ud}+V^2_{us}+V^2_{ub}$, and the most precise determination of the leading element $V_{ud}$ comes from nuclear superallowed $\beta$ transitions, through which nuclear physics involves.
Moreover, in order to extract the $V_{ud}$ value from these superallowed $\beta$ transitions, not only the experimental data but also several theoretical corrections are needed, including the isospin-symmetry-breaking corrections $\delta_c$.
In finite nuclei, the isospin symmetry is broken mainly due to the Coulomb interaction.

By using the self-consistent RHF+RPA approach, we performed a systematic calculation on the isospin-symmetry-breaking corrections $\delta_c$ for nine different superallowed $\beta$ transitions \cite{Liang2009}.
It has been found that the $\delta_c$ values are not sensitive to specific nuclear effective interactions but to the proper treatment of the Coulomb part.
The Coulomb exchange term \cite{Gu2013} plays an important role in this specific topic.
The $V_{ud}$ value was then deduced by combining with the experimental data and theoretical radiative corrections at that time.

\begin{figure}\centering
    \includegraphics[width=.45\textwidth]{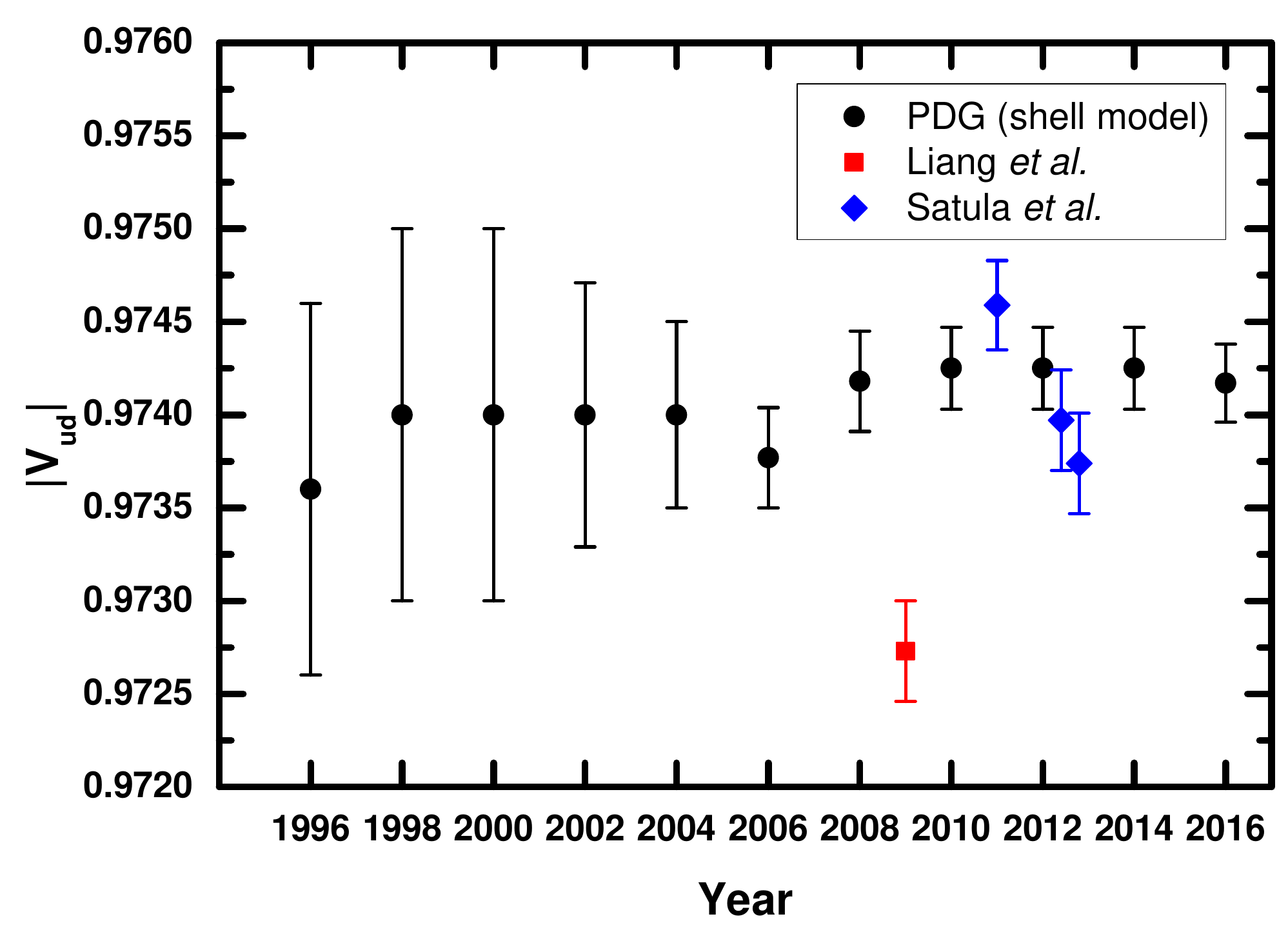}
    \hspace{1em}
    \includegraphics[width=.462\textwidth]{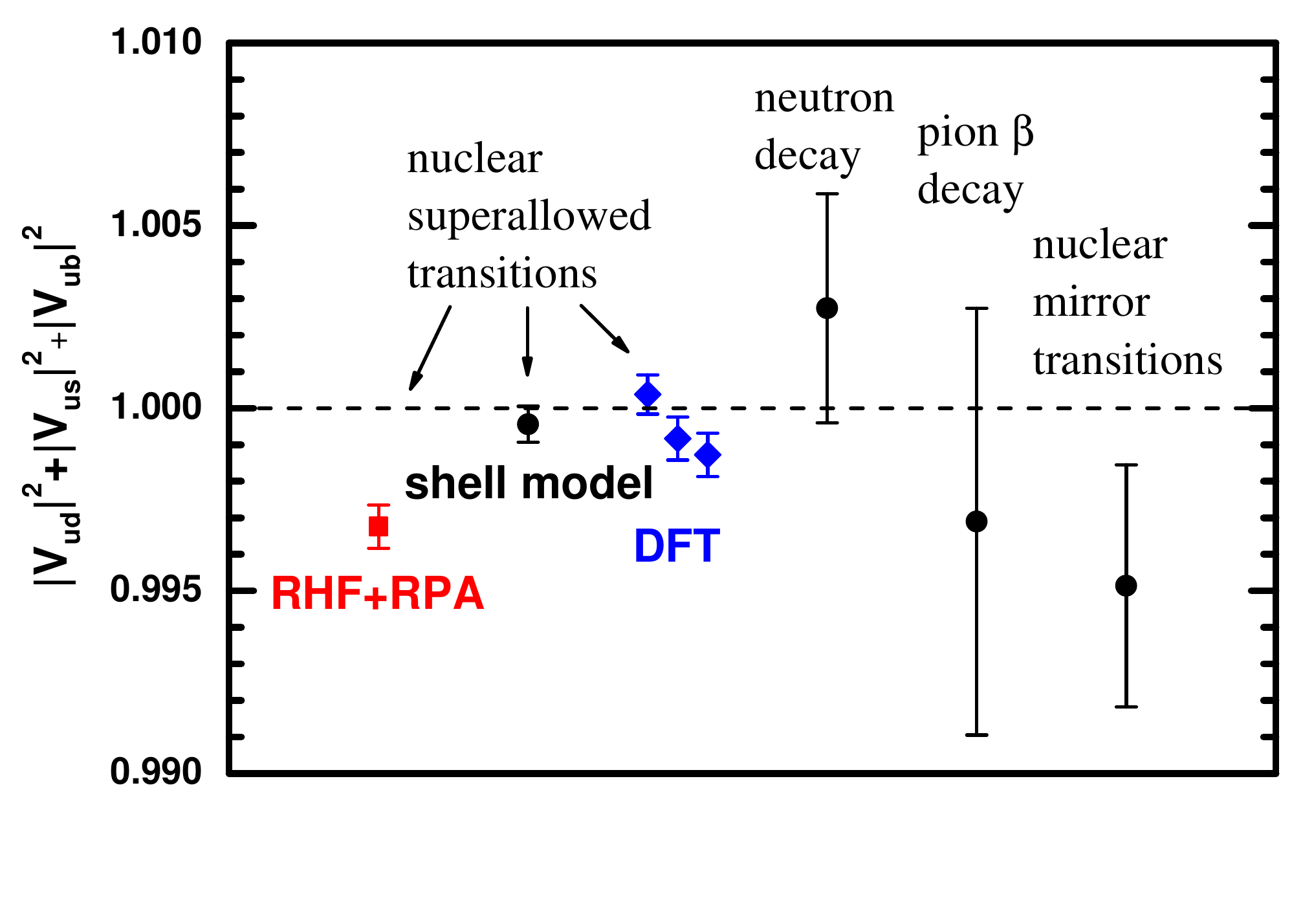}
    \caption{(Left) Values of $V_{ud}$ as a function of year.
    (Right) Square sums of $V^2_{ud}+V^2_{us}+V^2_{ub}$ obtained by different methods.}
    \label{Fig2}
\end{figure}

In the left panel of Fig.~\ref{Fig2}, I show the values of $V_{ud}$ as a function of year.
For over two decades, PDG takes the isospin-symmetry-breaking corrections $\delta_c$ from the shell-model calculations.
However, our RHF+RPA result \cite{Liang2009} as well as the results obtained by the projected DFT by Satu\l{}a \textit{et al.} \cite{Satula2011, Satula2012} are substantially different from the shell-model results.
This discrepancy has also been pointed out by PDG \cite{PDG}.

Combining the $V_{us}$ and $V_{ub}$ values shown in PDG2016, the corresponding square sums of the first-row elements are shown in the right panel of Fig.~\ref{Fig2}.
It can be seen that the unitarity condition is satisfied within the error bar by using the $\delta_c$ values from the shell-model calculations, while this conclusion is, however, much less clear if one uses the $\delta_c$ values from the DFT calculations.
But it is important to remark that in the DFT scheme the same nuclear effective interaction is used globally for all different superallowed $\beta$ transitions.

At this moment, the error bars are still large in other independent methods for determining the $V_{ud}$ value.
In particular, the central value obtained from the neutron-decay measurements has been shifted up substantially, comparing with the value given several years ago, which is because a \textit{shorter} neutron lifetime was found.
Therefore, it is still an open question, at least to me, whether (or how) the CKM matrix satisfies the unitarity condition.
This gives us one more strong motivations for improving the nuclear density functional theories.

\section{Relativistic Brueckner-Hartree-Fock theory for finite nuclei}

One of the ongoing projects concerns the tensor effects of nuclear interaction.
It is known that such tensor effects are crucial, in particular, for the properties of exotic nuclei, such as the new magic numbers.
Within the scheme of covariant DFT, on one hand, some fingerprints of tensor interaction can been seen on the two-proton separation energies \cite{Long2007} or the shell evolutions \cite{Tarpanov2008, Moreno-Torres2010}.
However, on the other hand, the tensor interaction is totally unwelcome if it is treated as a free parameter to fit the nuclear masses \cite{Lalazissis2009}.
In order to solve this puzzle, one may look carefully at the region on the nuclear chart where both \textit{ab initio} methods and DFT work.
In such cases, the \textit{ab initio} calculations could serve as a strong guild line for the development of DFT.

Another strong motivation comes from the latest progress in the lattice QCD simulations.
As mentioned above, to link the covariant DFT to more fundamental theories is one of the dreams, which may be indeed not very far away.
For example, during the last five years, our colleagues in HAL QCD collaboration have improved the simulations from a small box to a quite large box, from a heavy pion mass to almost the physical pion mass \cite{HAL}.
Therefore, we should also be ready for such progress from our side.

In the relativistic scheme, we chose the relativistic Brueckner Hartree-Fock (RBHF) theory as a benchmark.
In particular, we are developing the RBHF theory for finite nuclei.
For the first time, the Bethe-Goldstone equation was solved self-consistently in the two-body frame and in a pure relativistic scheme \cite{Shen2016}.

\begin{figure}\centering
    \includegraphics[width=.465\textwidth]{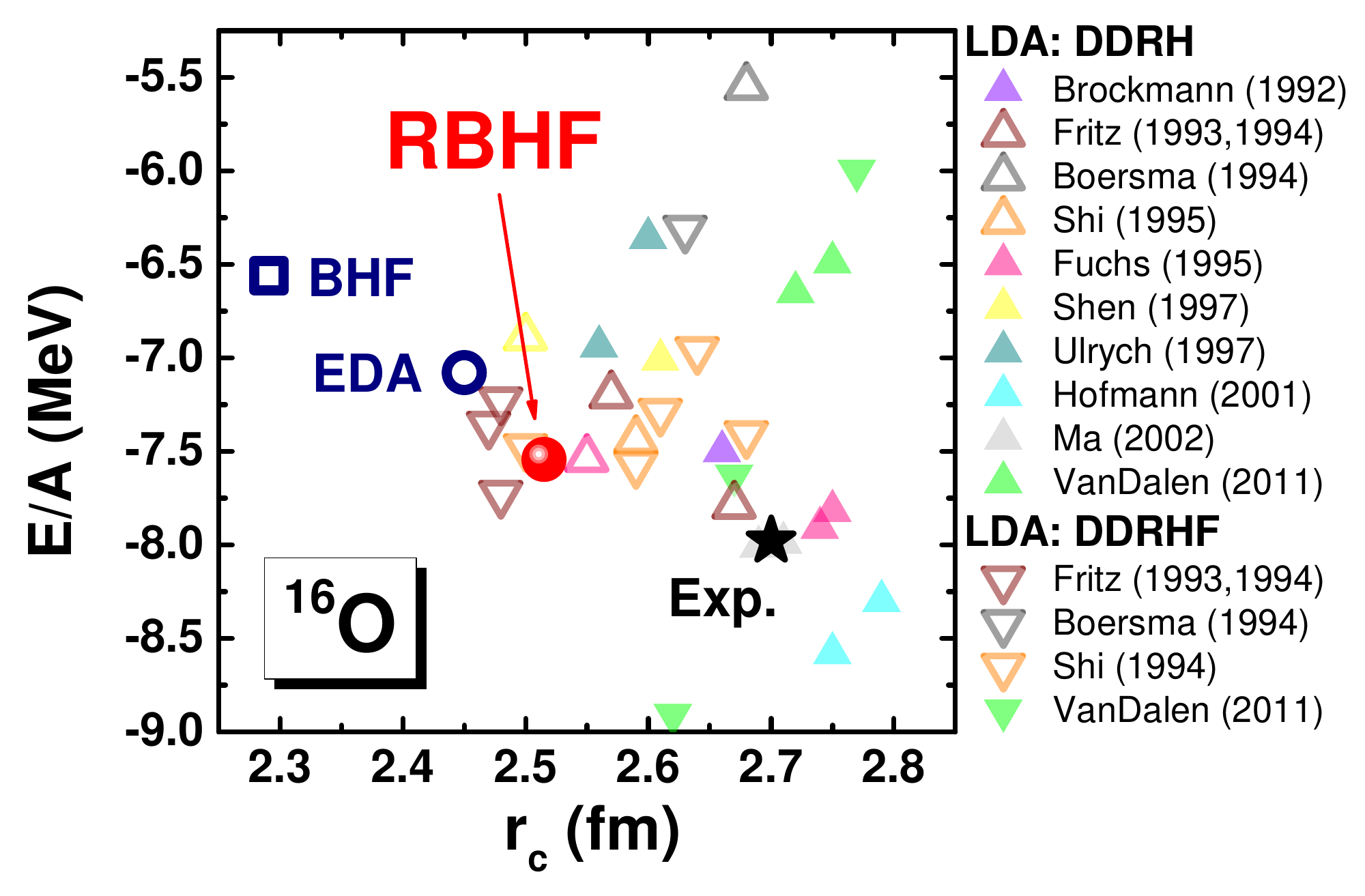}
    \hspace{1em}
    \includegraphics[width=.45\textwidth]{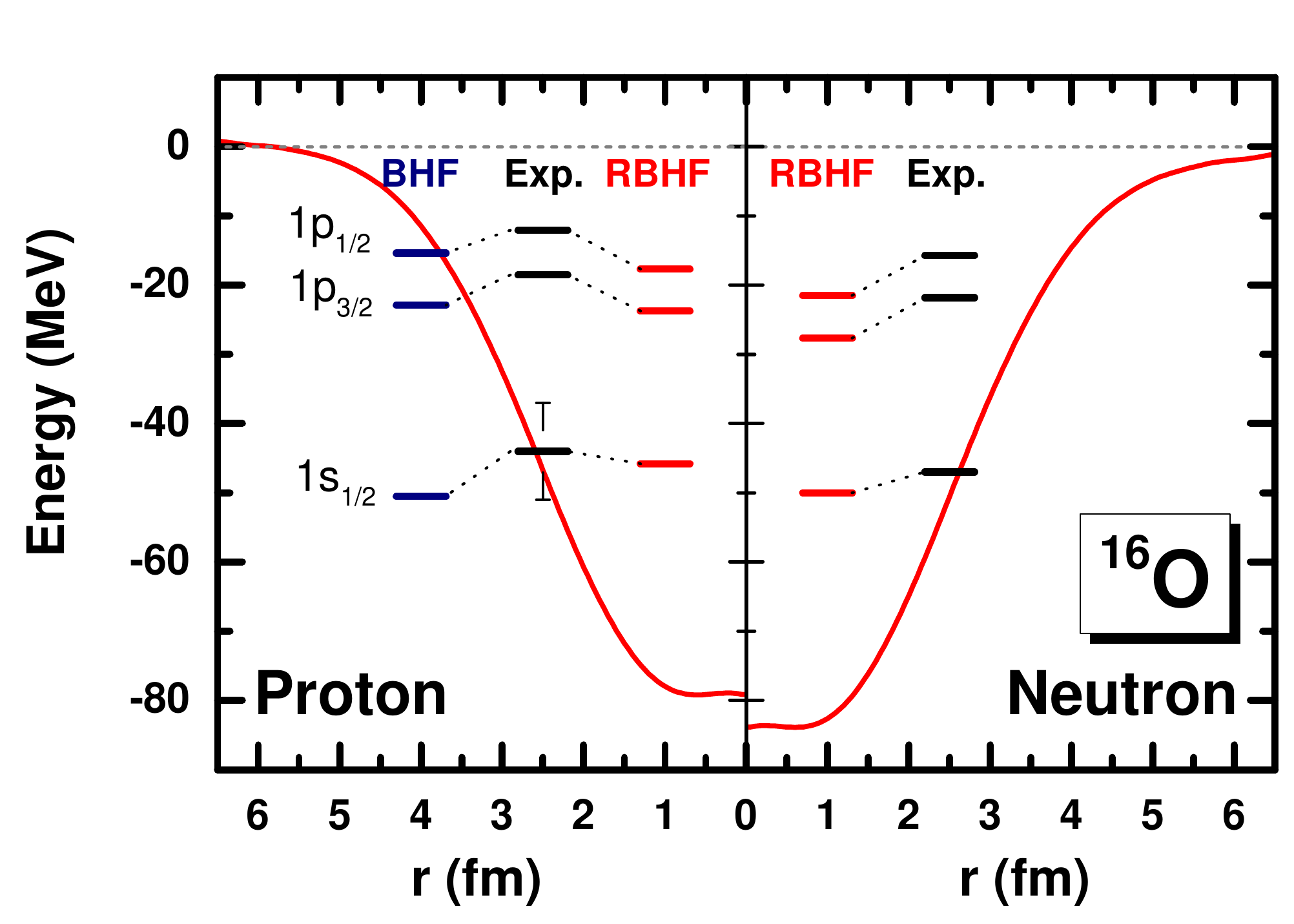}
    \caption{Energy per particle and charge radius (left) and single-particle spectra (right) of $^{16}$O by the RBHF theory with Bonn A interaction, comparing to the EDA (open circle), LDA (triangles), and non-relativistic (open square) calculations and the experimental data.
    Taken from Refs.~\cite{Shen2016, Shen2017}.}
    \label{Fig3}
\end{figure}

Figure~\ref{Fig3} shows the results of RBHF calculations for $^{16}$O with Bonn A interaction.
First of all, the non-relativistic BHF calculation without the three-body force is away from the data.
For many years, different kinds of approximation were introduced, in order to meet the computational power at that time.
Here shows the result from the effective density approximation (EDA), which is a kind of hybrid model between the non-relativistic and relativistic schemes.
Meanwhile, from the early 1990s, different kinds of local density approximation (LDA) were introduced by different groups.
Nevertheless, within the same framework by using the same interaction for the same nucleus, as shown with the triangles, the results are found to be very different from each other.
By avoiding these approximations, the present result would sever as a solid benchmark for various EDA/LDA calculations.
Furthermore, it is remarkable that the spin-orbit splitting is well reproduced from the bare nucleon-nucleon interaction, as we expect that the spin degree of freedom can be taken into account in a consistent way by using the Dirac equation in the relativistic framework.
The capability of RBHF calculations now extends to $^{48}$Ca in this ongoing project \cite{Shen2017}.

\section{Perspectives}

\begin{figure}\centering
    \includegraphics[width=.45\textwidth]{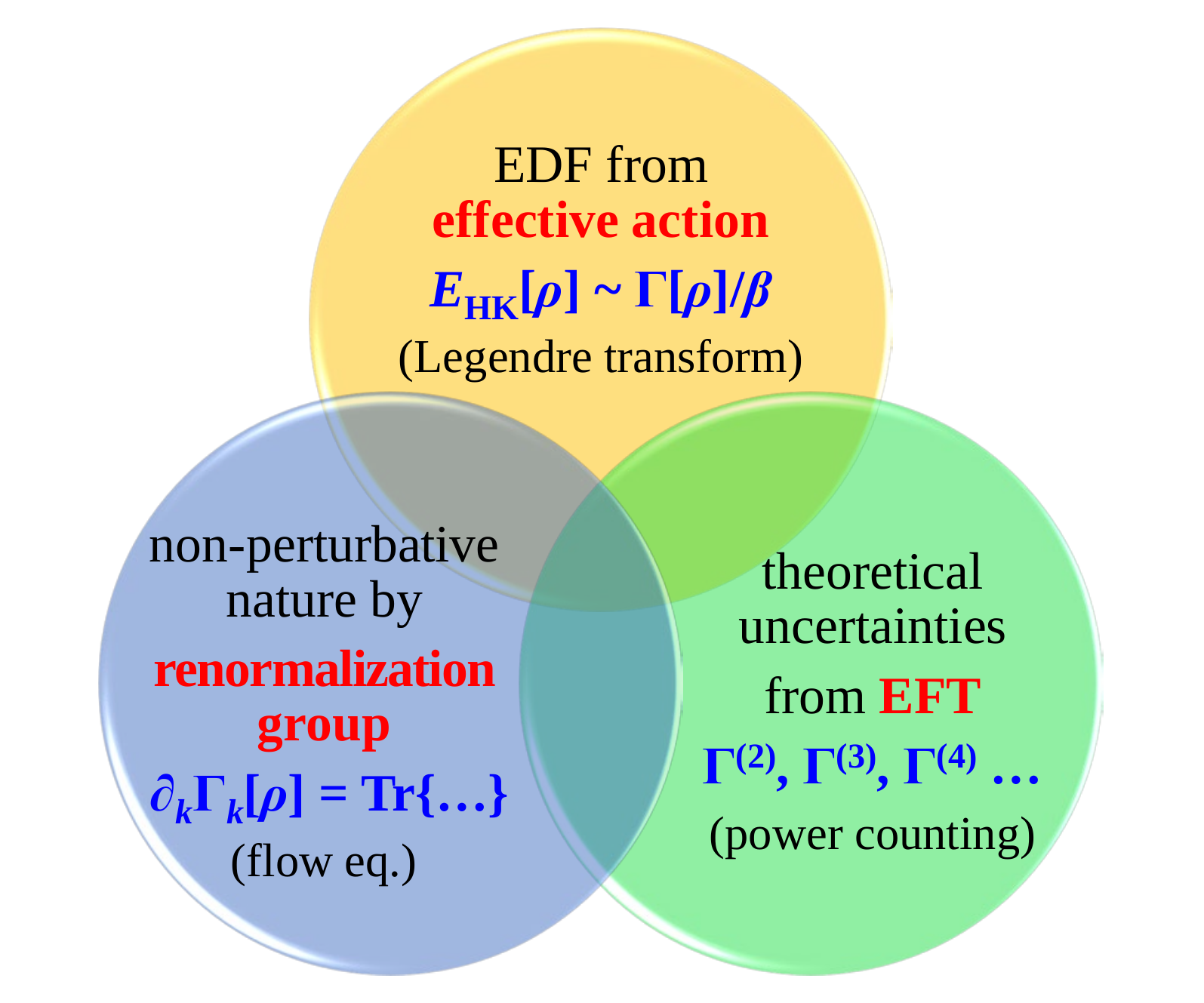}
    \caption{A dream for nuclear DFT oriented by quantum field theories.}
    \label{Fig4}
\end{figure}

In this special occasion, let me try to close this talk with a dream for another decade.
One of our dreams is to develop a new-generation nuclear DFT oriented by quantum field theories.
As illustrated in Fig.~\ref{Fig4}, we are considering in this scheme, (i) the energy density functional is derived from the effective action with Legendre transform, (ii) the non-perturbative nature of nuclear force is handle by the renormalization group with flow equations, and (iii) the theoretical uncertainties come with the idea of effective field theory with proper power counting (e.g., cf. Refs.~\cite{Schwenk2004, Kutzelnigg2006, Drut2010, Braun2012, Metzner2012, Drews2017}).

Within such a scheme, indeed we are able to share lots of knowledge and techniques with various fields, such as (lattice) QCD, hadron physics, cold atom physics, condensed matter, quantum chemistry, and so on.
Moreover, in the coming years, we will get a strong support from the developments of supercomputers.
Last but not least, with the construction and upgrades of new-generation experimental facilities all around the world, we will be able to benchmark and challenge each other.

\acknowledgments

It is my great honor to receive the IUPAP Young Scientist Prize in INPC2016.
I would like to thank the Commission on Nuclear Physics (C12) for the encouragement.
I wish to give my deepest gratitude to my Ph.D. supervisors Professor J. Meng, Professor P. Schuck, and Professor N. Van Giai, my postdoc host Professor T. Nakatsukasa, and my present division head Professor T. Hatsuda.
With you I have wonderful experiences and sweet memories in Peking University, Universit\'e Paris-Sud XI, and RIKEN.
I would like to appreciate Professor K. Fukushima and Professor T. Otsuka for bringing me into the graduate sub-course of Nuclear Theory in the University of Tokyo, so that I am able to work with excellent students.
I would like to express my gratitude to my colleagues and collaborators for all of the interesting results we have got and for the ongoing projects, in particular to M. Anguiano, Y. Chen, G. Col\`{o}, V. De Donno, T. Doi, Z. Fang, M. Grasso, H.Q. Gu, J.Y. Guo, K. Hagino, C. Ishizuka, Y. Kim, F.Q. Li, J. Li, L.L. Li, N. Li, Z. Li, Z.P. Li, Z.X. Li, Y. Lim, Q. Liu, W.H. Long, F. Minato, M. Moreno-Torres, P. Naidon, T. Nik\v{s}i\'{c}, Y.F. Niu, Z.M. Niu, N. Paar, J. Peng, B. Qi, P. Ring, X. Roca-Maza, H. Sagawa, M. Sasano, S.H. Shen, M. Shi, L.S. Song, C. Stoyanov, K. Sugawara-Tanabe, B.H. Sun, Y. Tanimura, Y. Tanizaki, D. Tarpanov, H. Toki, T. Uesaka, D. Vretenar, S. Wanajo, S.Y. Wang, J.X. Wei, X.D. Xu, J.M. Yao, L.F. Yu, S.Q. Zhang, S.S. Zhang, W. Zhang, Y. Zhang, P.W. Zhao, and S.G. Zhou.
\textit{To Jun and Junbin!}

\end{document}